\makeatletter
\declare@file@substitution{revtex4-1.cls}{revtex4-2.cls}
\makeatother
\documentclass[preprint2]{aastex631}

\newcommand\editone[1]{\textcolor{black}{#1}}
	\newcommand{\tbn}{$\theta_{Bn}$}

	\newcommand{\noopsort}[1]{}
	
	\def\physrep{\ref@jnl{Phys.~Rep.}}   % Physics Reports

\begin{document}
		
		\title{Observation of a Fully-formed Forward--Reverse Shock Pair Due to the Interaction Between Two Coronal Mass Ejections at 0.5 au}
		
		\correspondingauthor{D.\ Trotta}
		\email{d.trotta@imperial.ac.uk}
		
		\author[0000-0002-0608-8897]{Domenico Trotta}
		\affiliation{The Blackett Laboratory, Department of Physics, Imperial College London, London SW7 2AZ, UK}
		
		\author[0000-0003-1589-6711]{Andrew P. Dimmock}
		\affiliation{Swedish Institute of Space Physics, 751 21 Uppsala, Sweden}
		
		\author[0000-0001-7171-0673]{Xochitl Blanco-Cano}
		\affiliation{Departamento de Ciencias Espaciales, Instituto de Geofísica, Universidad Nacional autónoma de México, Ciudad Universitaria, 04150 Ciudad de México, Mexico}
		
		\author[0000-0003-2701-0375]{Robert J. Forsyth}
		\affiliation{The Blackett Laboratory, Department of Physics, Imperial College London, London SW7 2AZ, UK}
		
		\author[0000-0002-3039-1255]{Heli Hietala}
		\affiliation{Department of Physics and Astronomy, Queen Mary University of London, London E1 4NS, UK}
		
		\author[0000-0001-6308-1715]{Na{\"i}s Fargette}
		\affiliation{The Blackett Laboratory, Department of Physics, Imperial College London, London SW7 2AZ, UK}
		
		\author[0000-0002-7653-9147]{Andrea Larosa}
		\affiliation{Istituto per la Scienza e Tecnologia dei Plasmi (ISTP), Consiglio Nazionale delle Ricerche, I-70126 Bari, Italy}
		
		\author[0000-0002-1890-6156]{No\'e Lugaz}
		\affiliation{Space Science Center, University of New Hampshire, Durham, NH 03824, USA}
		
		\author[0000-0001-6590-3479]{Erika Palmerio}
		\affiliation{Predictive Science Inc., San Diego, CA 92121, USA}
		
		\author[0000-0002-4921-4208]{Simon W. Good}
		\affiliation{Department of Physics, University of Helsinki, FI-00014 Helsinki, Finland}
		
		\author[0000-0003-2495-8881]{Juska E. Soljento}
		\affiliation{Department of Physics, University of Helsinki, FI-00014 Helsinki, Finland}
		
		\author[0000-0002-4489-8073]{Emilia K. J. Kilpua}
		\affiliation{Department of Physics, University of Helsinki, FI-00014 Helsinki, Finland}
		
		\author[0000-0002-9707-3147]{Emiliya Yordanova}
		\affiliation{Swedish Institute of Space Physics, 751 21 Uppsala, Sweden}
		
		\author[0000-0002-7638-1706]{Oreste Pezzi}
		\affiliation{Istituto per la Scienza e Tecnologia dei Plasmi (ISTP), Consiglio Nazionale delle Ricerche, I-70126 Bari, Italy}
		
		\author[0000-0003-3623-4928]{Georgios Nicolaou}
		\affiliation{Department of Space and Climate Physics, Mullard Space Science Laboratory, University College London, Dorking RH5 6NT, UK}
		
		\author[0000-0002-7572-4690]{Timothy S. Horbury}
		\affiliation{The Blackett Laboratory, Department of Physics, Imperial College London, London SW7 2AZ, UK}
		
		\author[0000-0002-0074-4048]{Rami Vainio}
		\affiliation{Department of Physics and Astronomy, University of Turku, FI-20014 Turku, Finland}
		
		\author[0000-0003-3903-4649]{Nina Dresing}
		\affiliation{Department of Physics and Astronomy, University of Turku, FI-20014 Turku, Finland}
		
		\author[0000-0002-5982-4667]{Christopher J. Owen}
		\affiliation{Department of Space and Climate Physics, Mullard Space Science Laboratory, University College London, Dorking RH5 6NT, UK}
		
		\author[0000-0002-7388-173X]{Robert F. Wimmer-Schweingruber}
		\affiliation{Institute of Experimental and Applied Physics, Kiel University, D-24118 Kiel, Germany}

		
		%% argument for \collaboration is the collaboration identifier. authors are
		
		\begin{abstract}
			
			We report direct observations of a fast magnetosonic forward--reverse shock pair observed by Solar Orbiter on March 8, 2022 at the short heliocentric distance of 0.5~au.
			The structure, sharing some features with fully-formed stream interaction regions (SIRs), is due to the 
			interaction between two successive coronal mass ejections (CMEs), never previously observed to give rise 
			to a forward--reverse shock pair. The scenario is supported by remote observations from extreme ultra-violet cameras and coronographs, where two candidate eruptions compatible with the in-situ signatures have been found. 
			In the interaction region, we find enhanced energetic particle activity, strong 
			non-radial flow deflections, and evidence of magnetic reconnection. At 1~au, well radially-aligned \textit{Wind} 
			observations reveal a complex event, with characteristic observational signatures of both SIR and CME--CME 
			interaction, thus demonstrating the importance of investigating the complex dynamics 
			governing solar eruptive phenomena.
		\end{abstract}
		
	\keywords{Sun: coronal mass ejections (CMEs) --- Sun: heliosphere --- (Sun:) solar wind}
	
	
	\section{Introduction} \label{sec:intro}
	
	The Sun is an active star, responsible for generating a highly complex and dynamic 
	environment in its surroundings, namely the heliosphere.
	The solar activity and global field structures manifest themselves in a broad range of temporal and spatial scales in the heliosphere.
	
	The most common global structures that strongly influence the heliosphere 
	are stream interaction regions (SIRs) and 
	coronal mass ejections (CMEs). Understanding their origin and 
	propagation is pivotal for a broad range of 
	applications. Such phenomena play an important role in the production of 
	energetic particles and in the overall heliosphere energetics \citep{Rice2003}. SIRs and 
	CMEs also represent major drivers of the Sun--Earth interaction, making their
	investigation crucial from a space weather perspective \citep{Temmer2021}.

	SIRs form when the fast solar wind emerging from solar coronal holes 
		interacts with  the slow solar wind upstream of it \citep[see][for a review]{Richardson2018}.
		SIRs are crucial for planetary space weather \citep{Zhang2007}, and are the main source of 
		heliospheric suprathermal particles at solar minimum.
		SIRs are characterised by a region of compressed plasma, bounded by a 
		pair of forward--reverse pressure waves, which can steepen into forward--reverse shocks, 
		travelling away from and towards the Sun in the solar wind 
		rest frame, respectively \citep{Belcher1971}. 
		At 1~au, it was shown that less than 1\% of the SIRs are associated with 
		forward--reverse shock pairs \citep{Jian2006}, with even fewer observations in the inner 
		heliosphere below 1~au, as shown by earlier \textit{Helios} observations~\citep{Schwenn1996}.

	CMEs are the largest eruptive events from the Sun, defined as an observable 
	change in the coronal structure and an outward motion away from the solar 
	atmosphere \citep{Schwenn1996}. 
	They propagate at large heliocentric distances, and their rate is proportional to solar 
	activity. CMEs are excellent systems of 
	energy conversion, from the release of magnetic energy at 
	their origin to the shock-mediated conversion bulk flow energy into heat and energetic
	particles during their propagation \citep{Chen2011}. 
	In situ, CMEs show characteristic observable signatures and are 
	often separated in a forward shock (not always present), a compressed sheath region, 
	and magnetic ejecta \citep[][]{Kilpua2017}. 
	
	Increased solar activity introduces the opportunity to study the interaction between multiple 
	CMEs, which may happen in a variety of ways with different in-situ
	signatures \citep[see][for a review]{Lugaz2017}. Multiple-CME events 
	may lead to intense geomagnetic 
	storms \citep{Scolini2020} and extremely 
	intense solar energetic particle (SEP) 
	events \citep{Zhuang2020}.
	Interacting CMEs are the object of
	flourishing scientific debate, establishing their role in
	heliospheric energetics \citep[][]{Lugaz2014, Palmerio2021b}.
	Forward--reverse shock pairs due to transient disturbances and CMEs were also reported at 1~au 
	using early International Sun-Earth Explorer (ISEE) observations \citep{Gosling1988}.

	The Sun is approaching the maximum activity of solar cycle 25, and novel datasets are 
	now available, due to the ground-breaking Parker Solar Probe \citep[PSP;][]{Fox2016} 
	and Solar Orbiter \citep{Muller2020} missions. Thus, a novel observational window for 
	solar eruptive phenomena has opened \citep[e.g.,][]{Dresing2023}.

	In this work, we exploit this new window by reporting, for the first time, a 
	fully formed forward--reverse shock pair driven by two interacting CMEs at short heliocentric 
	distances. 
	The shock pair is observed by Solar Orbiter as close to the Sun as 0.5~au on
	March 8, 2022. {The Sun was particularly active when the event took place, 
		and we identified two candidate CMEs originating on March 6 and 7 from the active
		regions (AR) 12957 and 12958, by combining solar disk, coronograph and heliospheric imagery from
		the Solar Terrestrial Relations
		Observatory Ahead \citep[STEREO-A;][]{Kaiser2008}. These observations were complemented with 
		near-Earth remote observations from  the Solar Dynamics Observatory  \citep[SDO;][]{Pesnell2012} and 
		the Solar and Heliospheric Observatory \citep[SOHO;][]{Domingo1995}. Finally, we study the evolution 
		of the shock pair using direct observations from the radially well-aligned \textit{Wind} \citep{Ogilvie1997} spacecraft at 1~au.} We describe the data 
	products used in Section~\ref{sec:data}, while the results are presented 
	in Section~\ref{sec:results}, and the conclusions reported in Section~\ref{sec:conclusions}.
	
	\section{Data} \label{sec:data}
	
	For remote-sensing observations, we use STEREO-A's Sun Earth Connection Coronal 
	and Heliospheric Investigation \citep[SECCHI;][]{Howard2008} suite, with focus on 
	the {Extreme Ultra-Violet Imager (EUVI) observing the solar disk, the} COR2 
	coronagraph, imaging the solar atmosphere up to 15 $R_{\odot}$, 
	and the Heliospheric Imager (HI) cameras, observing the heliosphere in the elongation 
		range $4^{\circ}$--$88.7^{\circ}$. We complement these data with near-Earth 
		observations from the Solar Dynamics Observatory \citep[SDO;][]{Pesnell2012}, imaging the solar 
		disk, and the Solar and Heliospheric Observatory \citep[SOHO;][]{Domingo1995}, imaging the 
		solar corona.
	
	%+++++++Figure 1 -- Large scale overview++++++++++++++++
	\begin{figure*}[p]
		\includegraphics[width=\textwidth]{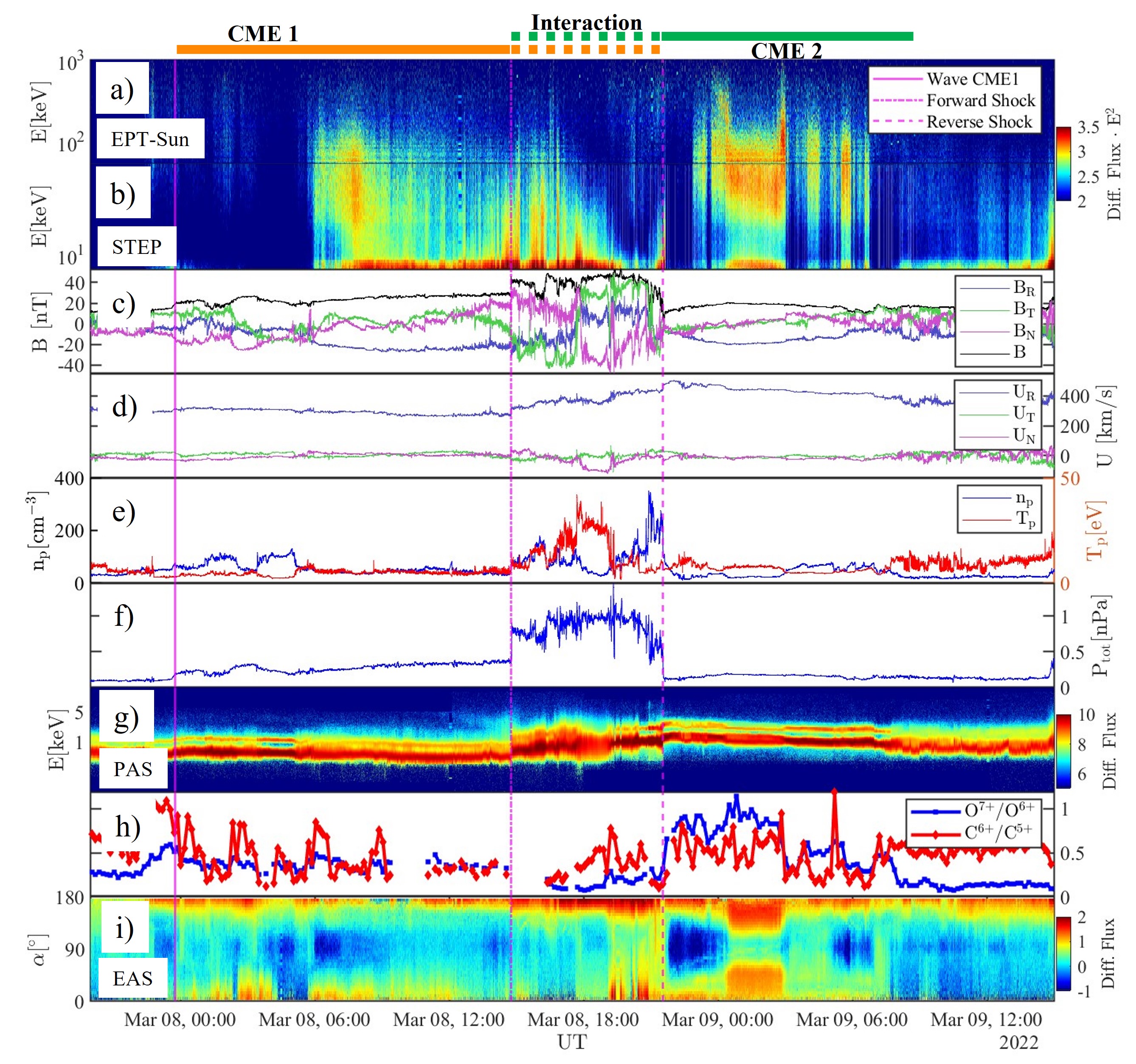}
		\caption{Summary of Solar Orbiter observations. a--b): Energetic ions differential fluxes (in $\mathrm{E^2 \cdot cm^{-2} s^{-1} sr^{-1}~MeV}$) as measured by EPD's Sun-directed Electron Proton Telescope (EPT, a) and Supra Thermal Electron Proton sensor (STEP, b). c) MAG normal mode magnetic field magnitude and components in \editone{spacecraft-centred Radial--Tangential--Normal (RTN) coordinates~\citep{franz2002}}. d-e) Proton bulk flow speed, proton density and temperature as measured by SWA-PAS (Proton Alpha Sensor). f) Plasma total pressure ($\mathrm{P_{tot}} = \mathrm{n_p k_B T_p} + \frac{\mathrm{B}^2}{2 \mu_0}$ where $\mathrm{k_B}$ and $\mathrm{\mu_0}$ are the Boltzmann constant and the vacuum magnetic permeability). g) One-dimensional energy flux (in $\mathrm{cm^{-2} s^{-1}~eV}$) measured by PAS. h) Element abundance ratios measured by the SWA Heavy Ion Sensor (HIS). i) Integrated pitch angle distributions for electrons with energies larger than 100 eV as measured by SWA-Electron Analyser Sensor (EAS) (in $\mathrm{cm^{-2} s^{-1} eV}$). The continuous, dashed--dotted and dashed lines show the times at which Solar Orbiter crosses the CME1 wave, the forward and reverse shock, respectively.}
		\label{fig:fig_overview}
	\end{figure*}
	%+++++++++++++++++++++++++++++++
	
	%------------------------Table of shock parameters-------------------------
	\begin{table*}
		\centering
		\caption{Shock times and parameters inferred from Solar Orbiter direct observations. The parameters shown are (left to right): shock normal vector, \tbn, magnetic compression ratio $\rm{r}_B$, gas compression ratio $\rm{r}$, shock speed $v_{\mathrm{sh}}$, upstream plasma beta $\beta_{\mathrm{up}}$, fast magnetosonic and Alfv\'enic Mach numbers ($\rm{M_{\rm{fms}}}$ and $\rm{M_{\rm{A}}}$, respectively).}
		\label{tab:tab_event}
		\begin{tabular}{lccccccccr} % four columns, alignment for each
			\hline
			Shock  &  Time [UT] & $\langle \hat{\mathrm{n}}_{\mathrm{RTN}} \rangle$ &$\langle \theta_{Bn}\rangle$ [$^\circ$] & $\langle \rm{r}_B \rangle$ & $\langle \rm{r} \rangle$  & $\langle v_{\rm{sh}} \rangle \rm [km/s]$ &  $\beta_{\mathrm{up}}$& $\rm{M_{\rm{fms}}}$ & $\rm{M_{\rm{A}}}$ \\
			\hline
			Forward & 08-Mar-2022 14:04:26 & [0.95 -0.11 0.30] & 59 & 1.5 & 2 & 367 & 0.09 & 1.2 & 1.1 \\
			Reverse & 08-Mar-2022 21:33:01 & [-0.94 0.15 0.32] & 69 & 2   & 2.1 & -373 & 3.1 & 2 & 3.8 \\
			\hline
		\end{tabular}
	\end{table*}
	%-----------------------------------------------------------
	At Solar Orbiter, we use the flux-gate magnetometer \citep[MAG;][]{Horbury2020}.
	Ion moments and suprathermal electron pitch angle distributions and composition are from 
	the Solar Wind Analyser suite \citep[SWA;][]{Owen2020}. 
	Energetic particles have been measured by 
	the Energetic Particle Detector \citep[EPD;][]{RodriguezPacheco2020}.
	
	From \textit{Wind}, we use the Magnetic Field
	Investigation \citep[MFI;][]{Lepping1995} measurements. {For \textit{Wind} plasma 
		measurements, we use data from the Solar Wind Experiment\citep[SWE;]{Ogilvie1995}, which we 
		checked against data from the Three-Dimensional Plasma and 
		Energetic Particle Investigation \citep[3DP;][]{Lin1995} instrument
		for ion moments.}

	\section{Results} \label{sec:results}
	
	%+++++++Figure 2 -- Zoom on interaction region ++++++++++++++++
	\begin{figure*}
		\includegraphics[width=\textwidth]{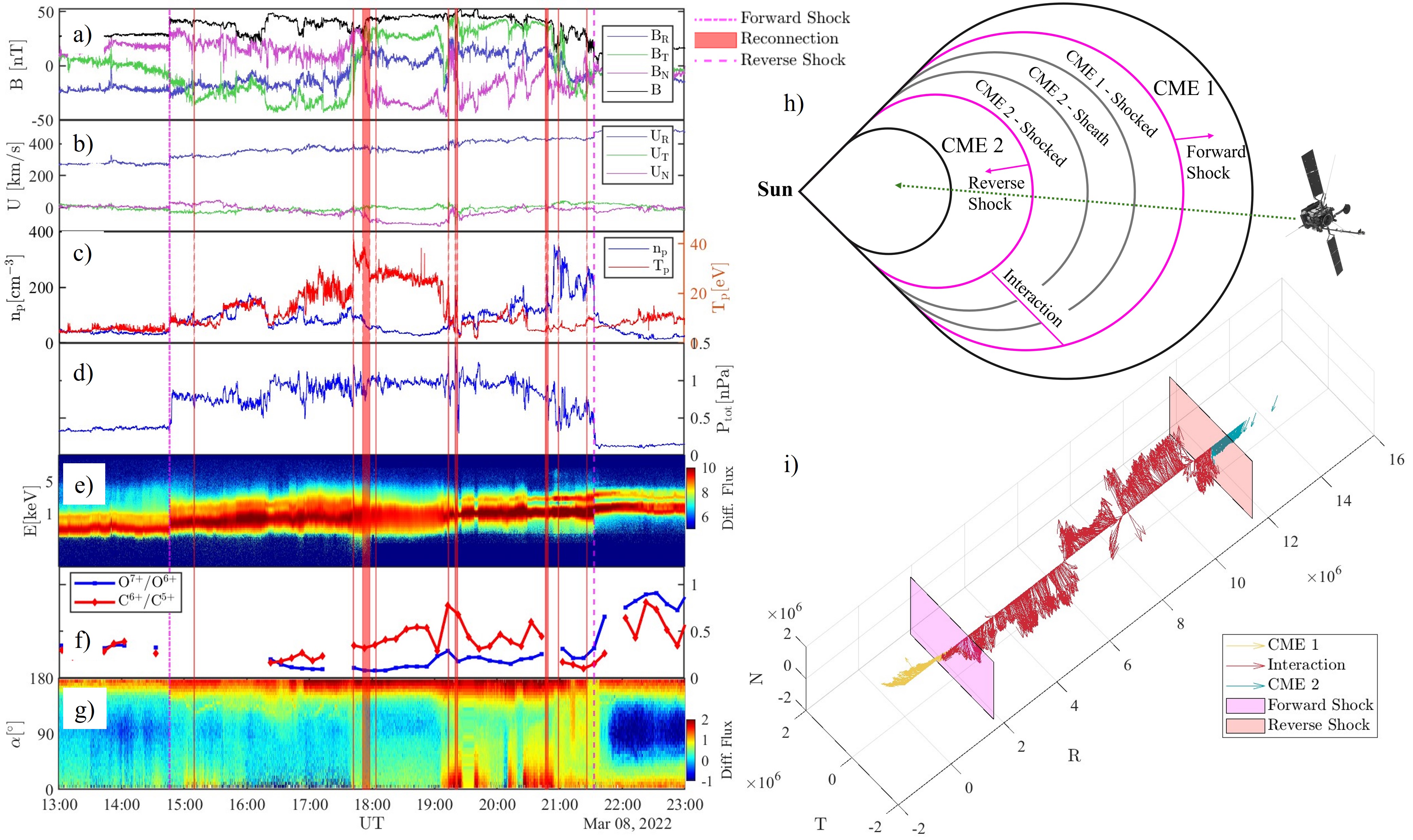}
		\caption{a-g: Zoom on the interaction region as in Figure~\ref{fig:fig_overview} without the energetic particles spectrogram. h: Simplified sketch representing the event {assuming head-on interaction} with the identified areas within the interaction and the Solar Orbiter trajectory (spacecraft model: esa.com\footnote{\url{https://www.esa.int/Science_Exploration/Space_Science/Solar_Orbiter}}). i: Three-dimensional plot of magnetic field vectors in RTN for the event. Yellow, red and blue arrows are measurements taken in CME1, interaction and CME2 regions, respectively. The magenta/orange planes represent the forward-reverse shock pair.}
		\label{fig:fig_interaction}
	\end{figure*}
	%+++++++++++++++++++++++++++++++
	
	%\subsection{Solar Orbiter Observations}\label{subsec:observations}
	On March 8, 2022, Solar Orbiter crossed a forward--reverse 
	shock pair at 0.49~au (Figure~\ref{fig:fig_overview}). The time separating the two 
	shocks is 6 hours and 47 minutes.% (about 10$^{10}$m assuming a bulk flow 
	%speed of 400 km/s). 
	
	{Figure~\ref{fig:fig_overview} shows an overview of the event, 
		with the forward--reverse shock pair highlighted by the dashed magenta lines. The interaction 
		region between the shocks is associated with magnetic field 
		compression, two subsequent increases in the bulk flow speed, and enhanced total plasma pressure 
		(see dashed and dotte-dashed lines in Figure~\ref{fig:fig_overview}c, d, f). While the structure is reminiscent of a fully-formed SIR, we note 
		unexpected profiles of some parameters, such as the hot, low density interior.
		Furthermore, forward--reverse shock pairs at such short heliocentric distance are not consistent with any 
		previous observations associated with SIR evolution, generally yielding forward--reverse shock pairs at 1~au and 
		beyond \citep[see, e.g.,][]{Richardson2018}}. 
	
	A closer analysis of the event reveals that the forward--reverse shock pair is not driven by an SIR, where fast 
	and slow solar wind interact. Rather, the forward--reverse shock pair is driven by the interaction 
	between two CMEs (CME1 and CME2 in chronological order, see Figure~\ref{fig:fig_overview})  with different 
	propagation speeds (about 290 and 450 km/s, respectively). 
	
	This is readily seen by the presence of several clear indicators of CME material \citep[see][]{Zurbuchen2016}
	both before and after the interaction region, including the smooth magnetic field rotations upstream/downstream 
	of the interaction, the enhanced O$^{7+}$/O$^{6+}$ ratios and the bi-directional pitch angle distributions of 
	suprathermal electrons (Figure~\ref{fig:fig_overview}c, h, i). Figure~\ref{fig:fig_overview}  shows that the CME1 
	is both slower (Figure~\ref{fig:fig_overview}d) and possibly magnetically less well-connected to the Sun (less 
	clear bidirectional electron signature).
	
	The start of CME1, on March 7, 2022, 7:23:46~UT, magenta line in Figure~\ref{fig:fig_overview} has not steepened into a shock due to the slow CME1 propagation speed.
	Downstream of the CME1 compressive wave, we observe a change of 
	parameters (around 6:00~UT on March 8), marking the start of enhanced energetic particle
	fluxes within the CME1 ejecta (Figure~\ref{fig:fig_overview}a, b).
	Protons with energies of up to 7~MeV were found, irregularly distributed 
	within both the CME1 and CME2 ejecta and in the interaction region. 
	This behaviour may depend on the intrinsic complexity 
	of the environment measured. Furthermore complexity may be due to further injection of energetic particles at 
	the Sun, where we identified a type III radio 
	burst around 4:30~UT (not shown here).

	The shock parameters, computed using the SerPyShock package~\citep[][]{Trotta2022b},
	systematically changing the upstream/downstream averaging 
	windows from a few seconds to 2 minutes, are summarized in Table~\ref{tab:tab_event}. 
	The forward shock is oblique (\tbn\,$\sim 60^\circ$), with small Mach numbers, while the reverse shock appears 
	more perpendicular and stronger, compatible with previous studies of SIR
	shocks \citep{Kilpua2015_shocks}. Starting at about 13:00~UT, we report enhanced particle fluxes upstream of and well-connected to the forward shock, readily seen in at high energies in the spectrogram in Figure~\ref{fig:fig_overview}g. This extended particle foreshock propagating in CME1 and thereby producing foreshock waves will be object of further study addressing the shocks' small-scale behaviour.

	%+++++++Figure 3 -- remote observations and orbital configuration++++++++++++++++
	\begin{figure*}[p]
		\includegraphics[width=\textwidth]{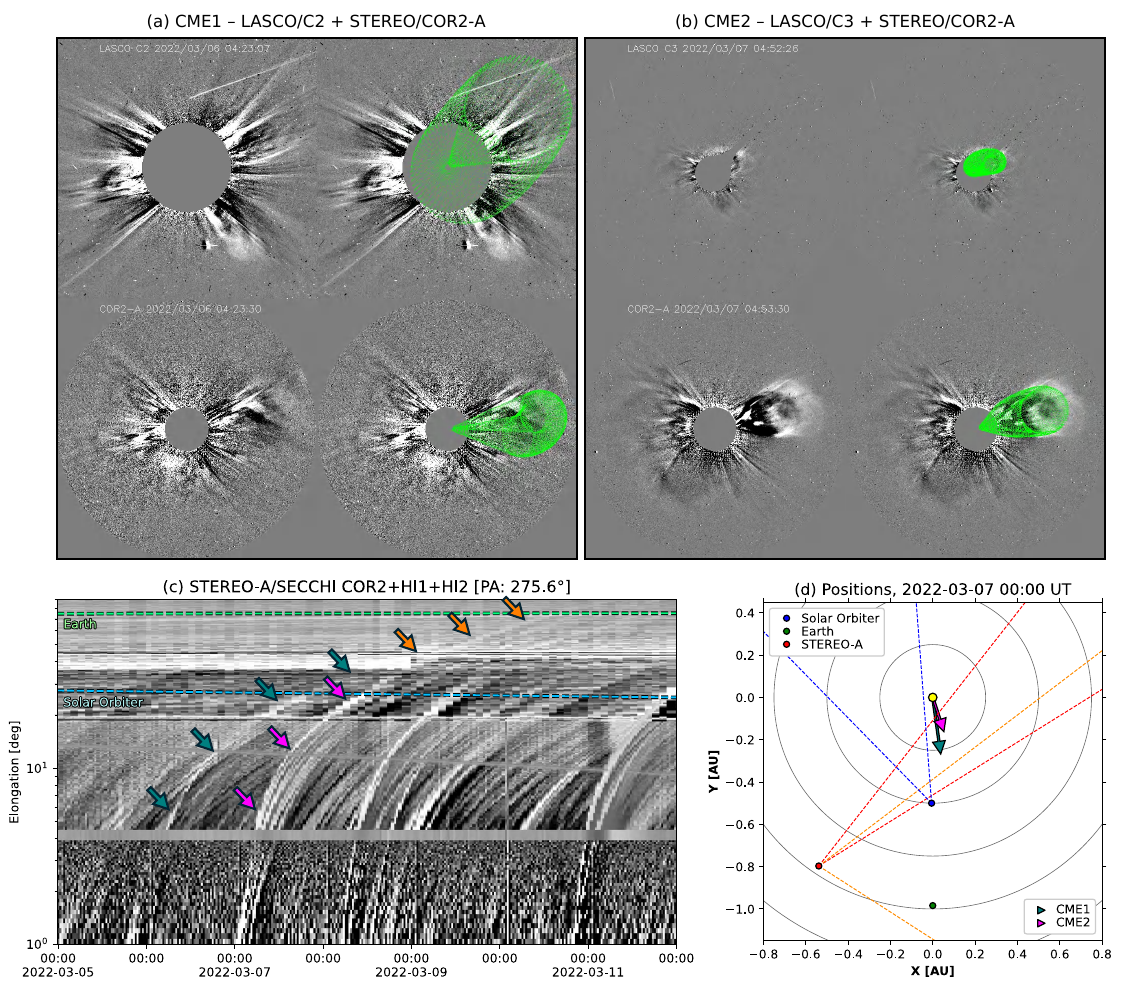}
		\caption{{Overview of some available remote-sensing observations of the 2022 March CME--CME interaction event. (a--b) Coronagraph observations of (a) CME1 and (b) CME2 from the (top) SOHO and (bottom) STEREO-A viewpoints. The rightmost panels show the GCS wireframe projected onto each plane-of-sky view. (c) Time--elongation map built using data from the COR2, HI1, and HI2 cameras on board STEREO-A. The tracks of CME1 and CME2 are indicated with teal and magenta arrows, respectively, and the combined track after interaction is shown by orange arrows. The time-dependent elongation angles of Solar Orbiter and Earth are highlighted via light blue and green dashed lines, respectively. (d) Orbit plot showing the relative positions of Solar Orbiter, Earth, and STEREO-A during March 7, 2022. The propagation directions of the two CMEs according to the GCS results are shown with arrows, while the dashed lines indicate the fields of view of the SoloHI (blue), HI1-A (red), and HI2-A (orange) heliospheric imagers.}}
		\label{fig:fig_orbit}
	\end{figure*}
	%+++++++++++++++++++++++++++++++

	We focus on the interaction region properties, showing a zoom of the Solar Orbiter measurements 
	together with a simplified sketch of the event in Figure~\ref{fig:fig_interaction}. {Note that the sketch assumes head-on interaction of the events, which may not be the case as argued below.}
	The trailing part of the interaction region is characterised by higher plasma densities, lower temperatures, 
	and higher elemental abundances (Figure~\ref{fig:fig_interaction}f) than the leading portion. 
	The pitch angle distributions show that magnetic connectivity changes in the end portion of the interaction 
	region from field lines connected to the Sun at one end to those having both ends connected. 
	These observations emphasise that the interaction region consists of plasma from two different 
	sources. 
	
	The interaction region shows sub-structuring,
	with irregular behaviour in many measured quantities (Figure~\ref{fig:fig_interaction} left).
	We suggest that this is due to the spacecraft probing, in rapid succession, the material at the 
	end of CME1 and material in the CME2 sheath and cloud (Figure~\ref{fig:fig_interaction}h), 
	as particularly evident in the plasma signatures (Figure~\ref{fig:fig_interaction}c, e). Plasma belonging to different regions may mix due to reconnection, discussed below.
	
	The leading part of the interaction region is characterised by hot plasma, strongly processed 
	by the forward shock.  {Progressing through the interaction, we observe 
		abrupt changes in magnetic field direction, in association 
		with strong transverse flow deflections (at 18:00 and 19:05~UT).These correspond 
		to plasma being deflected away
		from the radial direction in the interaction between the two events, 
		which may happen at an oblique angle, as further indicated by remote observations and discussed below}. We interpret this region as the interface between the two CMEs.
	
	As done in observations of
	planar magnetic structures in the solar wind \citep[e.g.,][]{Nakagawa1989} and in CME-driven sheath regions \citep{Palmerio2016},
	we applied a Minimum Variance Analysis (MVA) to 
	the magnetic field in the interaction region. 
	In the interval from the immediate downstream of the forward shock to the reverse shock, the 
	intermediate-to-minimum eigenvalue ratio of the MVA matrix is large ($\lambda_2/\lambda_3 \sim 8$). 
	This implies the existence of a well-defined minimum variance direction. Projecting the magnetic field 
	components in the MVA frame highlights the change at 
	the interface region at 18:00~UT mentioned above (not shown here). Strong changes
	at 19:00~UT are also found both with the MVA and magnetic reconnection diagnostics (see below), 
	indicating that there may be more than one interface crossing.
	
	Further characterization was performed, searching for 
	magnetic reconnection signatures, crucial for
	mixing plasmas efficiently \citep[e.g.,][]{Russell1990}. 
	We used the magnetic reconnection method 
	successfully applied to Solar Orbiter data in \citet{Fargette2023}. Orange shaded regions 
	in Figure~\ref{fig:fig_interaction} (left) correspond to reconnection exhaust crossings. 
	It is readily seen  
	that the interaction region undergoes strong reconnection activity, very long-lasting 
	around 18:00~UT, corresponding to the previously identified CME--CME interface and corroborating the
	interpretation of complex mixing of CMEs.
	
	Finally, in Figure~\ref{fig:fig_interaction}i 
	we show a three--dimensional plot of the magnetic field vectors as measured by Solar Orbiter 
	in the CME1, interaction region, and CME2 intervals, 
	with the forward--reverse shock
	pair represented as the magenta/orange planes, respectively. The interface between the two
	CMEs can be clearly seen in the 
	sharp change of direction of the magnetic field.
	
	The spacecraft orbital configuration during the event makes it possible to get unique insights 
	about the evolution of this novel interaction structure.
	{By combining solar disk, coronagraph, and heliospheric imagery from STEREO-A and near-Earth spacecraft}, 
	we identified two candidate eruptions from the Sun, possibly the progenitors
	of the observed interaction event.
	{An overview of our findings is provided in Figure~\ref{fig:fig_orbit}. CME1 appears as a faint event in STEREO-A imagery and is not visible in SOHO data, while CME2 can be observed in both views (Figure~\ref{fig:fig_orbit}a, b). By performing reconstructions of both events using the Graduated Cylindrical Shell \citep[GCS;][]{Thernisien2011} model, we find a propagation direction of ($\theta$, $\phi$) = ($-2^{\circ}$, $8^{\circ}$) for CME1 and ($\theta$, $\phi$) = ($9^{\circ}$, $18^{\circ}$) for CME2 in Stonyhurst coordinates, with speeds of $\sim$500~km/s and $\sim$425~km/s, respectively ( summarized in Table~\ref{tab:tab_cmes}). Such parameters indicate that the interaction between the two CMEs is not perfectly head-on, consistent with the tangential flow deflections found by means of direct observations at both Solar Orbiter and \textit{Wind}, as discussed below. We remark that the reconstruction for CME1 is performed using the STEREO-A viewpoint only and is thus associated with larger uncertainties \citep[see e.g.][]{Verbeke2023}. We note that CME1 is only slightly faster than CME2 according to GCS results, which is likely to lead to CME2 catching up with CME1 due to solar wind preconditioning \citep[e.g.,][]{Temmer2017}. The source region of CME1 is a stealth CME \citep[see e.g.][]{Palmerio2021a} from active region (AR)~12957 around 00:00~UT on March 6 (mostly visible off the limb from STEREO-A), while CME2 is associated with a more energetic eruption from AR~12958 around 00:00~UT on March 7. We follow the interplanetary propagation of the two CMEs via time--elongation maps that employ STEREO/SECCHI data (Figure~\ref{fig:fig_orbit}c), where the two CME tracks are seen to converge (possible indication of merging) beyond Solar Orbiter's heliocentric distance (see Figure~\ref{fig:fig_orbit}d for the geometry of the STEREO/HI fields of view). We note that we do not find CME signatures in heliospheric images from SoloHI on board Solar Orbiter, consistent with the two CMEs originating from the western hemisphere (see Figure~\ref{fig:fig_orbit}d).}
	%These are displayed in Figure~\ref{fig:fig_orbit} a, b for CME1 and CME2, 
	%respectively, while Figure~\ref{fig:fig_orbit} c shows the spacecraft 
	%orbital configuration with the STEREO-A field of view (red cone) 
	%and a model CME eruption (rainbow cone) obtained using the propagation tool in %\citet{Rouillard2017}.
	While identification of the remote counterparts of CME1 and CME2 
	is not straightforward, {our analysis shows that} both candidates are compatible with {eruptions directed close to the Sun--Earth line} and with their arrival time at Solar Orbiter. 
	%They are also compatible with the observation
	%of CME1 being slow and not well connected to 
	%the Sun and CME2 being faster and still well-connected to the Sun.
	Such remote observations highlight how
	even ``faint'' solar eruptions can give rise to energetic events through complex interactions.
	%------------------------Table of CME parameters-------------------------
	\begin{table}
		\centering
		\caption{Summary of remote observations and GCS fit results for CME1 and CME2. The $\theta$ and $\phi$ angles are in Stonyhurst coordinates.}
		\label{tab:tab_cmes}
		\begin{tabular}{lcccr} % four columns, alignment for each
			\hline
			CME  &  Time [UT] & $\theta$ [$^\circ$] &$\phi$ [$^\circ$]& Speed [km/s] \\
			\hline
			1 & 06-Mar 00:00 & -2 & 8 & 500 \\
			2 & 07-Mar 00:00 & 9 & 18 & 425 \\
			\hline
		\end{tabular}
	\end{table}
	%-----------------------------------------------------------
	%+++++++Figure 4 -- Wind observations++++++++++++++++
	\begin{figure*}
		\includegraphics[width=\textwidth]{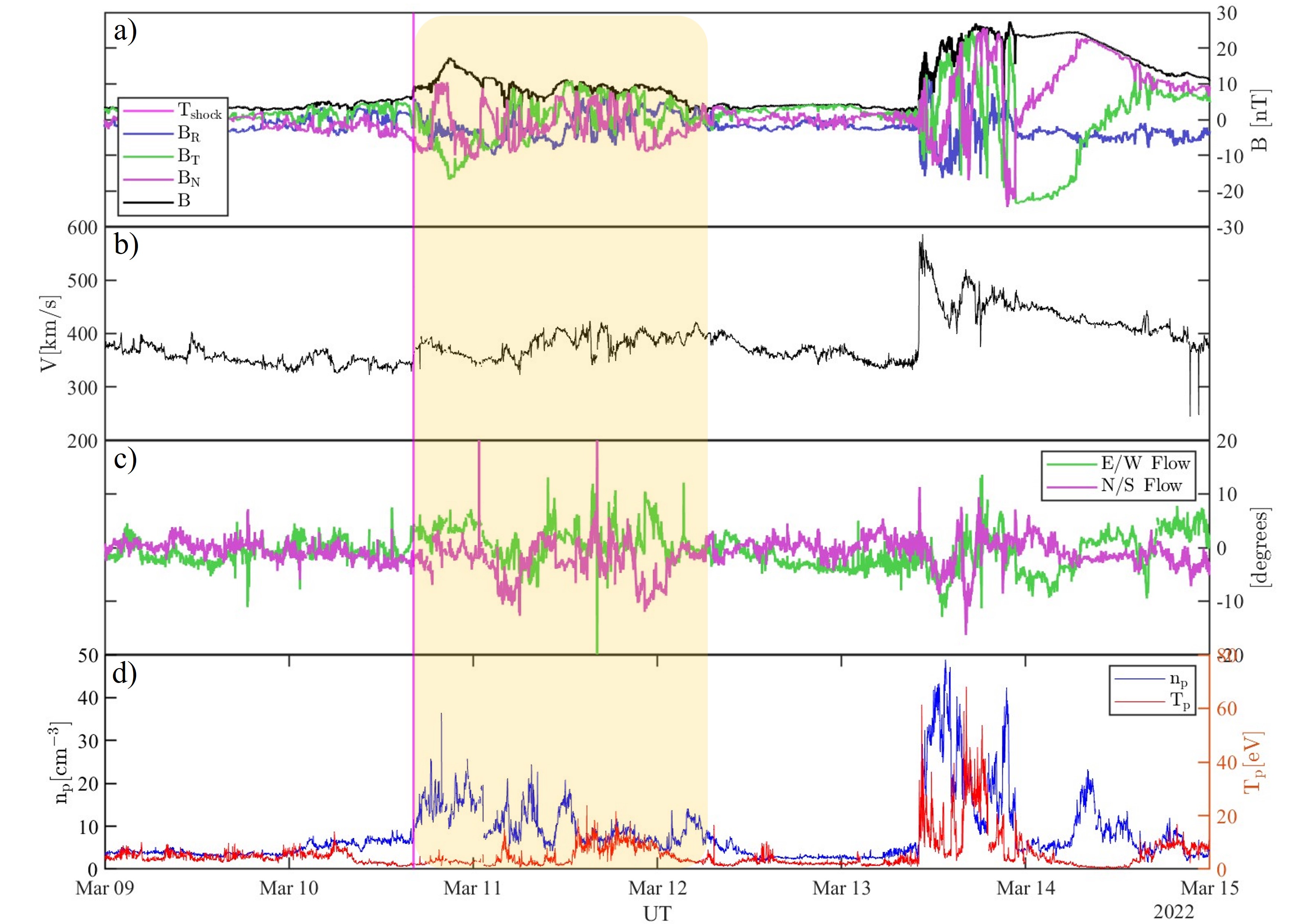}
		\caption{{ \textit{Wind} observations of the event at 1 au (shaded area). From top to bottom are displayed: magnetic field magnitude and components in RTN (a), proton bulk flow speed and tangential flows (b, c), protons density and temperature (d). The magenta line marks the forward shock crossing.}}
		\label{fig:fig_wind}
	\end{figure*}
	%+++++++++++++++++++++++++++++++
	
	%As already hinted in Figure~\ref{fig:fig_orbit}c, we exploit the unique opportunity to study the evolution of the interaction at 1~au. 
	During the event, the Solar Orbiter--Earth longitudinal separation was about $9^{\circ}$. We exploited the configuration and found an in-situ structure crossing the \textit{Wind} spacecraft at the Sun--Earth Lagrange L1 point
	at 13:00~UT on March 10, 2023, compatible with the Solar Orbiter event propagating at
	about 400 km\,s$^{-1}$ speed from 0.5 to 1~au.
	
	In Figure~\ref{fig:fig_wind}, it is readily noted that the forward--reverse shock
	pair is not present at \textit{Wind}.
	Only a fast-forward shock is observed at \textit{Wind} ahead of the whole structure, crossing the spacecraft on March 10 at 16:11:32~UT. The shock has a complex magnetic structure in both the upstream and downstream regions, which probably dominates its small-scale evolution features~\citep[see][]{Trotta2024}. A data gap in plasma measurements immediately downstream makes shock parameter estimation particularly difficult. Nevertheless, we estimate that the shock is oblique ($\theta_{Bn} \sim 55^\circ$) and the Alfvén Mach number is very low, close to 1.
	
	The event at \textit{Wind} is compatible with the complex ejecta resulting from the interaction of multiple CMEs, as reported in \citet{Lugaz2017}, where characteristics of the individual ``parent'' eruptions can no longer be discerned. {Transverse flow deflections are found at 1 au, also present in the 0.5 au observations. This indication of oblique interaction between the two CMEs is compatible with remote observations and the GCS fits discussed above (Figure~\ref{fig:fig_orbit}).}
	The structure at 1~au also has some features reminiscent of a SIR (e.g., magnetic compression), 
	but is missing the typical fast stream signature corresponding to the reverse pressure wave. We note that the solar wind speed is slower in the first portion of the complex ejecta than in the second, thus maintaining the general trends of the structure observed at Solar Orbiter. It is also possible that the 10$^\circ$ separation between the two locations was sufficient to measure different parts of the event between Solar Orbiter and the Sun--Earth L1 point. Therefore, these joint  observations highlight the transient nature of this novel interaction. 
	{The interaction has weakened from 0.5 to 1~au, as it can be seen in the 7-days overview plot in Figure~\ref{fig:fig_wind}, where the interaction appears as a very moderate event, quite common during solar maximum. This is particularly evident when compared  with another CME event crossing \textit{Wind} around 12:00 UT on March 13~\citep[see][for futher details]{Laker2024}. Such behaviour is in contrast with that expected from SIR driven forward--reverse shock pairs that tend to get stronger as they move to larger heliocentric distances \citep{Richardson2022}. }The event was not found 
	to cause any major space weather disturbance at Earth, consistent with the lack of periods with steady magnetic field orientations \citep{Dimmock2019_SW}.

	\section{Conclusions} \label{sec:conclusions}
	
	We reported direct observations of a fully--formed reverse--forward shock pair at the 
	very low heliocentric distance of 0.5~au. While such a shock pair is typically associated with an SIR, it was found to be originated from the interaction between 
	a fast and a slow CME. {To our knowledge, this is the first time that such an 
		observation is reported at such small heliocentric distances.}
	
	The CME--CME interaction drives a complex compression region, where the interface separates plasma from two different sources and is characterised by a high level of magnetic reconnection activity and several 
	irregularities in the measured plasma conditions. Such characterisation underlines the role of 
	this structure in creating favourable conditions efficient energy dissipation \citep{Richardson2018}.
	
	Energetic ions up to several MeV 
	were observed, with a strongly irregular behaviour, influenced by the 
	complex plasma environment, stimulating an advancement of knowledge for
	energetic particle behaviour in the heliosphere.
	On one hand, due to novel, 
	high time-energy resolution of energetic particles datasets \citep[see][]{Wimmer2021},
	we were able to link irregular particle 
	behaviour to the plasma irregularities
	(discontinuities, reconnection) present in the same region, a study in continuity with others
	using EPD in different environments \citep[e.g.][]{Trotta2023c}. On the other hand, it was shown 
	that a significant amount of high energy ($\sim$7~MeV) particles may be generated
	in the interaction between weak eruptive events, with important consequences
	for ongoing modelling efforts in SEP acceleration and propagation~\citep[e.g.][]{Ding2024}.
	
	The forward--reverse shock pair propagating in CME material also offers opportunity to 
	study shock micro-physics in unusual ambient parameters, as in the case of the forward shock
	exhibiting an extended particle foreshock despite the very low Mach number, probably due to 
	the low level of upstream magnetic field fluctuations of CME1 \citep[see][]{Trotta2021, Lario2022}. 
	Studying shock behaviour in this poorly explored parameter space is important
	for the astrophysical implications of this research, and will be the object of 
	further studies.
	
	{This study exploited the unique orbital configuration during the event, with two remote CME candidates
		identified using STEREO-A and near-Earth observers. These are compatible with CME1 being a faint eruption, 
		originating from AR 12957, then interacting with CME2, which is more energetic and originating from AR 12958, as shown by 
		the time-elongation maps in Figure~\ref{fig:fig_orbit}. Despite the large uncertainties involved, GCS fits yield CME propagation speeds compatible with this scenario. These observations highlighted the importance
		of connecting remote and direct observations, particularly due to CME1 being particularly faint and slow, yet giving rise to such an interesting event.}
	
	We also investigated the evolution of this structure at 1~au using the \textit{Wind}
	spacecraft, revealing a merged structure without forward--reverse shock pair and 
	mixed features between a CME and SIR event. {At 1~au, the structure became a moderate event, common around solar maximum, underlining that without an inner heliosphere upstream observer we would have little knowledge of its origins and evolution.}
	This is in contrast with SIR-related shock pairs, which get more intense with heliospheric distance. 
	The fact that such CME--CME-related shock pairs seem to weaken with heliocentric distance is compatible with the fact that they have not been identified previously, and with earlier simulation studies of interacting CMEs \citep{Lugaz2005}.
	
	To get further insights into the evolution of
	these transient, complex interactions has relevant implications to space weather 
	events \citep{Mostl2020}, and will be further investigated exploiting
	the extended spacecraft fleet orbiting the inner heliosphere.

	%% Also note that the akcnowlodgment environment does not support long amounts of text. If you have a lot of people and institutions to acknowledge, do not use this command. Instead, create a new \section{Acknowledgments}.
	\section*{Acknowledgments}
	%\section{acknowledgments}
	This study has received funding from the European Unions Horizon 2020 research and innovation programme under grant agreement no.\ 101004159 (SERPENTINE, \href{www.serpentine-h2020.eu}{www.serpentine-h2020.eu}).Views and opinions expressed are, however, those of the authors only and do not necessarily reflect those of the European Union or the European Research Council Executive Agency. Neither the European Union nor the granting authority can be held responsible for them. This work was supported by the UK Science and Technology Facilities Council (STFC) grant ST/W001071/1.
	Solar Orbiter magnetometer operations are funded by the UK Space Agency (grant ST/X002098/1).
	Solar Orbiter is a space mission of international collaboration
	between ESA and NASA, operated by ESA. Solar Orbiter Solar Wind
	Analyser (SWA) data are derived from scientific sensors which have been
	designed and created, and are operated under funding provided in
	numerous contracts from the UK Space Agency (UKSA), the UK Science and
	Technology Facilities Council (STFC), the Agenzia Spaziale Italiana
	(ASI), the Centre National d’Etudes Spatiales (CNES, France), the Centre
	National de la Recherche Scientifique (CNRS, France), the Czech
	contribution to the ESA PRODEX programme and NASA. Solar Orbiter SWA
	work at UCL/MSSL is currently funded under STFC grants ST/W001004/1 and
	ST/X/002152/1.
	The Energetic Particle Detector (EPD) on Solar Orbiter
	is supported by the Spanish Ministerio de Ciencia, Innovación y
	Universidades FEDER/MCIU/AEI Projects ESP2017-88436-R and PID2019-
	104863RB-I00/AEI/10.13039/501100011033 and the German space agency (DLR)
	under grant 50OT2002.
	X.B.-C.\ is supported by DGAPA-PAPIIT grant IN106724.
	H.H.\ is supported by the Royal
	Society University Research Fellowship URF\textbackslash R1\textbackslash 180671.
	N.F.\ is supported by the UKRI/STFC grant ST/W001071/1.
	A.L.\ and O.P.\ are supported by the PRIN 2022 project ``2022KL38BK - The ULtimate fate of TuRbulence from space to laboratory plAsmas (ULTRA)'' (Master CUP B53D23004850006) by the Italian Ministry of University and Research, funded under the National  Recovery and Resilience Plan (NRRP), Mission 4 – Component C2 – Investment 1.1, ``Fondo per il Programma Nazionale di  Ricerca e Progetti di Rilevante Interesse Nazionale (PRIN 2022)'' (PE9) by the European Union – NextGenerationEU. A.L.\ also acknowledges the support of the STFC Consolidated Grant ST/T00018X/1.
	E.P.\ acknowledges support from NASA's Heliophysics
	Guest Investigators-Open programme (grant no.\ 80NSSC23K0447).
	S.W.G.\ is supported by the Research Council of Finland (INERTUM, grant no.\ 346612).
	E.Y.\ is supported by Swedish National Space Agency (grant no.\ 192/20).
	N.D.\ acknowledges the support by the Research Council of Finland (SHOCKSEE, grant no.\ 346902). 
	
	T.S.H. is supported by STFC grant ST/W001071/1.
	\editone{Note that we used the RTN coordinate system throughout this work, routinely used for heliospheric missions. However, transformations to other systems of coordinates may be performed using the SPICE kernels available at \url{https://www.cosmos.esa.int/web/spice/solar_orbiter} for Solar Orbiter and at \url{https://naif.jpl.nasa.gov/naif/} for \textit{Wind} and STEREO.}

\end{document}